\documentclass[a4paper]{article}
\ifx\pdfoutput\undefined
\usepackage{graphicx}
\else
\usepackage[pdftex]{graphicx}
\fi

\usepackage{units}
\usepackage{balance} 

\usepackage{amsmath}
\usepackage{url}
\newcommand{\ie}{{\em i.e.}~}
\newcommand{\lFig}[1]{\label{fig:#1}}

\newcommand{\lEq}[1]{\label{eq:#1}}
\newcommand{\seeEq}[1]{Eq.~\ref{eq:#1}}
\newcommand{\seeFig}[1]{Fig.~\ref{fig:#1}}

\usepackage{subfigure}

\begin{document}
\ifx\pdfoutput\undefined
\graphicspath{{./figures/fig_eps/}}
\else
\graphicspath{{./figures/fig_pdf/}}
\fi
\title{Coding static natural images using spiking event times : do neurons cooperate?}
\author{Laurent Perrinet, %
Manuel Samuelides~\emph{Member,~IEEE}, %
Simon Thorpe
\thanks{L. Perrinet is with INCM/CNRS, France and the referring author (e-mail: \texttt{Laurent.Perrinet@incm.cnrs-mrs.fr}), M.~Samuelides with DTIM/ONERA, France and S.~Thorpe with \sc{CerCo}/CNRS}}
\markboth{IEEE Trans. on Neural Networks}{Perrinet \MakeLowercase{\textit{et al.}}: Coding static natural images using cooperating spiking event times}%
\date{}%
\maketitle
\begin{abstract}
To understand possible strategies of temporal spike coding in the central nervous system, we study functional neuromimetic models of visual processing for static images. We will first present the retinal model which was introduced by Van Rullen and Thorpe \cite{Vanrullen01} and which represents the multi-scale contrast values of the image using an orthonormal wavelet transform. These analog values activate a set of spiking neurons which each fire once to produce an asynchronous wave of spikes. %
According to this model, the image may be progressively reconstructed from this spike wave thanks to regularities in the statistics of the coefficients determined with natural images. Here, we study mathematically how the quality of information transmission carried by this temporal representation varies over time. In particular, we study how these regularities can be used to optimize information transmission by using a form of \textit{temporal cooperation} of neurons to code analog values. The original model used wavelet transforms that are close to orthogonal.
However, the selectivity of realistic neurons overlap, and we propose an extension of the previous model by adding a \textit{spatial cooperation} between filters. This model extends the previous scheme for arbitrary ---and possibly non-orthogonal--- representations of features in the images. In particular, we compared the performance of increasingly \textit{over-complete representations} in the retina. Results show that this algorithm provides an efficient spike coding strategy for low-level visual processing which may adapt to the complexity of the visual input.
\end{abstract}
\section{A dynamic representation of a static world}
\subsection{The spiking nature of the neural code}
Spikes reveal a paradox in our knowledge of the brain. Since their discovery, it is known that these brief and relatively intense peaks of the neuron's membrane potential are an almost universal feature of nervous systems. Using the terminology of signal processing, since they are mostly similar for all neurons, we may describe spikes as "all-or-none" events generated by a non-linear "explosive" mechanism at the neuronal membrane. Moreover, since the transmission of spikes between neurons through the synapses is highly reproducible \cite{Mainen95}, they are often regarded as the only mechanism for long-range inter-neuronal communication, the remaining signal in the membrane potential being "filtered out". This neuronal information, forming a spatio-temporal pattern of spikes, is then presumably "decoded" by the neurons's dendrites ---even along relatively long distances--- in a highly parallel fashion (neurons receive around $10^4$ synapses from other neurons). Finally, these successions of computations is one of the core features of the various cognitive processes that characterize living species. Therefore, spikes seem to provide a simple universal medium for inter-neuronal communication. But, paradoxically, it is still unclear how these spikes are interpreted, \ie what eventual "neural spiking code" could be used.\\%
In particular, there is little agreement about the representation of the information used by the spatio-temporal pattern of spikes. Following the pioneering work of Adrian \cite{Adrian28}, classical theories suggest that each neuron effectively integrates its inputs by computing a correlation with a previously learned pattern of synaptic weights and that this analog value is translated into changes in the frequency of spike firing. Since the \textit{perceptron} model \cite{Rosenblatt60}, these models  have become increasingly more complex and now form the very rich and powerful class of algorithms  used in Artificial Neural Networks. These algorithms have many links with other domains from mathematics to engineering and have been used to solve numerous problems which were intractable using methods from Artificial Intelligence. However, and in particular in the sub-class of feed-forward models (\ie where communication loops are avoided), remaining information potentially contained in the detailed spatio-temporal spike patterns of biological neurons is often ignored. Recently, much progress has been done in focusing on the temporal course of these complex neuronal systems to provide dynamic theories of brain functions, particularly by including feed-back loops \cite{Dauce98}. \\%
As is revealed by the complex architecture of visual processing in the cortex of primates, neurons interact through different dynamic pathways \cite{Bullier01} and a growing number of recent theories of neuronal coding take into account the precise \textit{latency of the spikes} by using dynamic models. These advances are inspired by recent neurophysiological studies which suggest that the "deviations" from the classical models may specifically carry an important part of the information, particularly over short epochs \cite{Panzeri99}, so that the actual complexity of the response of biological neuronal populations may in fact mirror more complex spatio-temporal relations. In fact, recent studies have revealed puzzling aspects of behavior: neurons that form populations may in some cases keep the same mean firing frequency but convey different informations by varying the overall degree of coherency in the population's spike pattern \cite{Zador98impact}. In this paper, we will explore how the cooperation in time and space of neurons may provide a more efficient transmission of the information over its temporal evolution, hence a strategy of \textit{dynamic spike coding}.%
\subsection{Constraints on a temporal spike code}%
One of the most important evolutionary constraints on the neural code is the need to convey the information in a fast and robust fashion. In particular, the results of psycho-physiological experiments which showed the rapidity of categorization in primates \cite{Thorpe96} have stimulated the search for dynamic neuronal models compatible with these temporal constraints. Therefore, we have developed new paradigms of neural coding in feed-forward neural networks that include latency and rank-order based coding \cite{Thorpe01b}.\\%
In contrast with detailed neurophysiological models where spikes are the result of a large number of differential equations describing the neuron's behavior, these algorithms focus on the relative latency of spikes to build a functional temporal code. To stress this difference, the equations governing neuronal behavior are reduced to the strict minimum, and we only use the first "transient" wave of spikes generated by a parallel neuronal layer : neurons fire only once and the information is encoded in the relative latency of spikes of this wave of single spikes. \\%
In particular, we analyzed the performance of different models of highly parallel networks of  asynchronous neurons. These neurons consist of simple elementary dynamic "voting" devices which cooperate to provide on demand a fast or robust decision. These ideas were implemented in the {\sc Spikenet} model \cite{Delorme99spikenet} and have proved to be very efficient for pattern recognition, mimicking the performance of biological processing. Thus, this direction of research proves that a more realistic model of a temporal spike code does not necessarily need to be more complex, but rather can exploit all the efficiency of the parallel and asynchronous structure of biological processing in the central nervous system. As a result, the rank-order coding scheme provides an original set of dynamic systems. They result in a wide class of novel and efficient algorithms that can be applied to signal processing.%
\subsection{Models of ultra-rapid image coding}
To study realistic models of temporal spike coding, we will therefore study here a model of spike coding in a functional framework. Due to its relative simplicity and to the extensive research on this part of the brain, we are especially concerned in this paper with visual processing in the retina. Specifically, we will be interested in how the information sensed by the photo-receptors could be efficiently encoded  by the spatio-temporal pattern of spikes generated by the ganglion cells which then project to the brain via the optic nerve. We therefore applied and studied models of spiking neurons to a simple model of information transmission in the optic nerve using temporal spike coding. \\%
Through this paper, we will use  natural static images as inputs. These are defined as images that are typical of those that occur in real life and include \textit{e.g.} outdoor scenes. We will analyze how the model retina reacts to static flashed images firstly to follow the protocol of Ultra-Rapid Categorization but also for the sake of simplicity. A more realistic temporal model would need to include more complex dynamics including adaptation and eye movements. Also, this will provide an illustration of how dynamic algorithms also apply to static stimuli.\\%
In this paper, we will first present the architecture of the model introduced by Van Rullen and Thorpe \cite{Vanrullen01}. This retinal model uses a precise wavelet-like transform based on the responses of ganglion cells to form a complete temporal spike code in the retina using an orthonormal representation. We will then evaluate the information transmission through the spikes and propose some alternatives to this model based on the statistics of natural images which improves the temporal cooperation of neurons in time. Then, we will propose an alternative scheme which complements this model by using lateral interactions and show how non-orthogonal representations may be used to build a generic neural code in the central nervous system. We will discuss the relative efficiency of this  in the retina for growing numbers of neurons. We finally propose a strategy for a temporal spike code using the \textit{spiking events} as a substrate for neural computation.%
\subsection{Methods of quantitative analysis}
To rate the quality of the reconstruction we will use (as in \cite{Vanrullen01}) the Mean-Squared Error (or \emph{MSE}) which measures the mean energy of the difference between the image $I$ and its reconstruction $I_\mathtt{rec}$ over the pixels $l \in \mathcal{I}$ where the image is defined:%
\begin{eqnarray}%
\mathtt{MSE}(I,I_\mathtt{rec}) = E[\|I-I_\mathtt{rec}\|_2^2] =  E[\sum\nolimits_{l \in \mathcal{I}} (I(l)-I_\mathtt{rec}(l))^2]\nonumber%
\end{eqnarray}%
Although an  image reconstruction  is biologically not very realistic, this criterion is particularly adapted for the retina if we consider it as the lowest level of the visual architecture and that it should provide a versatile representation. Moreover, the MSE provides ---under the assumption of Gaussian noise in a linear model of image construction and a uniform prior over the representation--- a direct measure of the log likelihood of image reconstruction knowing the initial image \cite{Rao99predictive}.\\%
Additionally, we will use Mutual Information from information theory. It is measured as the mean quantity of information that can be obtained about one image when the other is known (for a review, see \cite{Atick92retina}). It is therefore the sum of the entropies of the luminance of both images minus the coupled entropy:%
\begin{eqnarray}%
 \mathtt{MI}(I,I_\mathtt{rec}) &=  H[I]+ H[I_\mathtt{rec}]- H[I,I_\mathtt{rec}]\nonumber\\%
   =& \sum\nolimits_{l \in \mathcal{I}} P[I(l),I_\mathtt{rec}(l)].log(\frac{P[I(l),I_\mathtt{rec}(l)]}{P[I(l)].P[I_\mathtt{rec}(l)]})%
\nonumber%
\end{eqnarray}%
where $H$ represents the entropy and $P$ the probability. %
Throughout the paper the different algorithms are rated by evaluating these measures on a set of randomly chosen images defined over rectangular grids. These images were chosen in the  publicly available database of linearly calibrated natural images from Van Hateren as described in \cite{Hateren98}. These images were corrected using a $\gamma$ correction \cite{Poynton99} to assure the balance of luminance and mimic the analogical response of photo-receptors to luminosity, \ie to light intensity.%
\begin{figure*}[ht]%
\centering%
\subfigure[Architecture of the spiking model, (PhRs) linear integration part up to the Ganglion Cells and (GCs) non-linear spiking layer.]{\includegraphics[height=3.5cm]{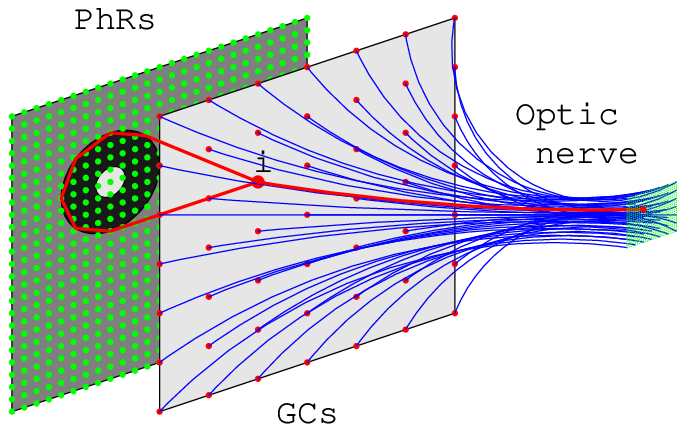}\label{fig:archi_a}}%
\hfill%
\subfigure[Sample input static gray-scale image: Lena at size $256\times2565$.]{\includegraphics[height=3.5cm]{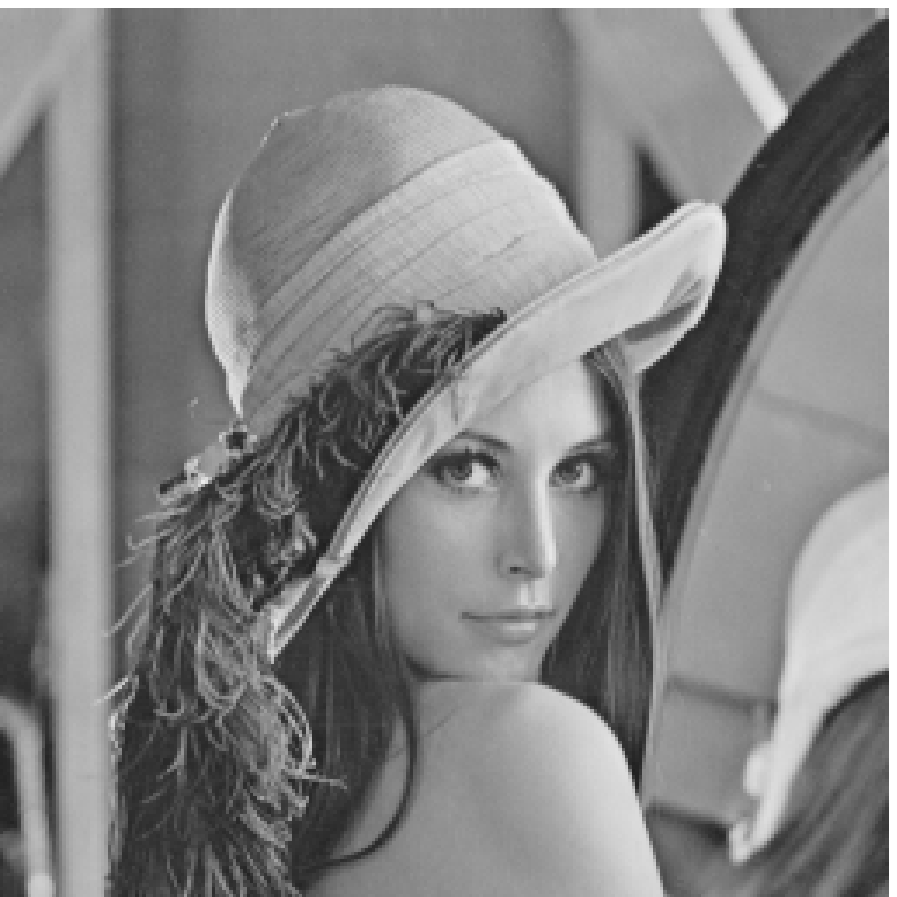}\label{fig:archi_b}}%
\hfill%
\subfigure[Montage of the multi-scale representation images, from lowest scale (left) to the highest (in spiral).]{\includegraphics[height=3.7cm]{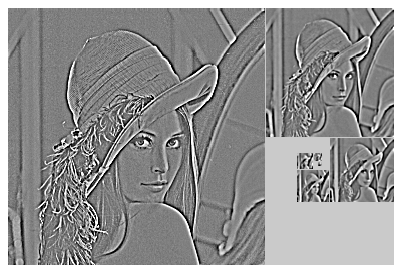}\label{fig:archi_c}}%
\caption{Architecture of the model retina. \ref{fig:archi_a} The model retina consist of two layers : an input layer corresponding to the photo-receptors (PhRs) which transmits the information analogically and linearly to the output layer corresponding to the ganglion cells (GCs). A sample GC neuron $i$ is highlighted in red, and we showed its center-ON receptive field soma and axon. These ganglion cells produce spikes along their axons (forming the optic nerve). \ref{fig:archi_b} The input of the model consist of static gray-scale images sampled over a rectangular grid (a white pixel corresponding to high luminance) which elicit the activity of the photo-receptors. \ref{fig:archi_c} We present here the activity of the ganglion cells which represents either positive (represented as white pixels) or negative (black) contrasts at different scales. We represented here for clarity the lowest scale (therefore at the highest frequency) with the left-most image and in a spiral the progressively down-sampled images of contrast values for progressively higher scales (\ie lower frequencies). Note that most values are gray, \ie of low activity : the distribution of coefficients is \textit{kurtotic}.}%
\lFig{archi}%
\end{figure*}%
\section{Temporal cooperation in an orthogonal wavelet architecture}%
To achieve a comparison with other models, we will define our model retina as the classical feed-forward two-layered neural network described in \cite{Vanrullen01}. It is characterized by a set of neurons, the ganglion cells (GCs), sensitive at different spatial scales to the local contrast of the image intensity detected at the input layer of the photo-receptors. These GCs then emit spikes for which we will propose a compact temporal spike code. These assumptions are unlike the biological retina which first has a hexagonal sampling but also has a higher concentration of neurons near the optical axis (the fovea), but are sufficient to describe as a first step a static retina.%
\subsection{Architecture of the linear retinal model: from light to multi-scale contrast representation}%
Based on  neurophysiological data \cite{Rodieck65}, we assume that this part of the architecture (and which drives the activity at the soma of the GCs) results from linearly filtering features of the luminance distribution. The dendrite of a neuron $i$ may thus be characterized by its weight vector $\phi_{i}$ over the image and to mimic neurophysiological constraints (total length of dendrites) and data, these functions are often localized on a receptive field of limited radius (see \seeFig{archi_a}). The activity at the soma of the neuron is the usual dot product :%
\[%
 C_{i} := <I,\phi_{i}> = \sum\nolimits_{l\in {\mathcal{R}_i} } I(l).\phi_{i}(l)%
\]%
where $I(l)$ is the luminance at pixel $l$ (see an example input image at  \seeFig{archi_b}) and ${\mathcal{R}_i}$ is here the receptive field of the neuron $i$.\\%
For the sake of simplicity and to apply this algorithm with standard computerized images, the photo-receptors and neurons are placed uniformly over rectangular grids. We also assume that the architecture is both translation independent (that is that neurons on a same scale are replicated over the different positions) and also scale invariant. %
 From these assumptions, we can define a single \textit{mother function} $\psi$ from which every filter can be  derived using  translation and scaling. If the activity is computed over all positions $l$ and scale $s$, the set of activity values constitute a continuous wavelet transform. A particular case of a discrete wavelet transform chosen by Van Rullen and Thorpe \cite{Vanrullen01} is to choose a \textit{dyadic} progression, \ie where filter radius and grid spacing both grow in a geometric fashion as powers of $2$. This choice is a particular down-sampling of the continuous wavelet transform which is particularly frequent in image processing (see \seeFig{archi_c}).\\%
As the architecture is defined, an important task is to choose an appropriate mother wavelet to detect contrasts in the image. As in \cite{Vanrullen01} and from \cite{Field94}, neurons are defined here according to their position $l_c$ and scale $s$ as dilated, translated and sampled {\em Mexican Hat} (or Difference Of Gaussian --- DOG) filters (see \cite[pp. 77]{Mallat98}, and \seeFig{dog}) as %
\begin{eqnarray}%
 \mathtt{DOG}_{\{s, l_c\}}(l) &=& 9.G_{\sigma(s)}(l-l_c)-G_{3.{\sigma(s)}}(l-l_c)\\%
\mbox{with }%
G_{\sigma(s)}(l)&=&\frac{1}{\sqrt{2\pi}.{\sigma(s)}}.\exp(-\frac{{\| l\|}^2}{2.{\sigma(s)}})%
\end{eqnarray}%
where we denote $G_{{\sigma(s)}}$ as the 2D Gaussian function of variance ${\sigma(s)}$ which itself depends on the scale $s$. %
\begin{figure}[h]
\centering
\subfigure[Image of a DOG filter contrast detector.]{%
\hskip .5cm%
\includegraphics[height=3cm]{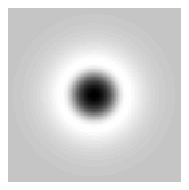}%
\hskip .5cm%
}\label{fig_dog_a}%
\subfigure[Slice of the retinal filter along the axis.]{%
\hskip .5cm%
\includegraphics[height=3cm]{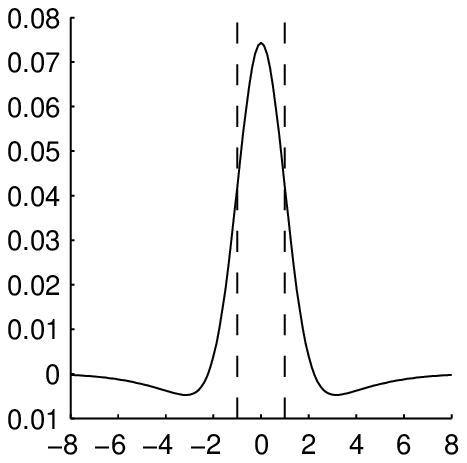}%
\hskip .5cm%
 \label{fig_dog_b}}%
\caption{\textit{Linear filters in the retina}. The DOG filter in \ref{fig_dog_a} is a center-OFF surround-ON contrast detector.  As a unit measure, the vertical striped lines in \ref{fig_dog_b} represent the variance of the narrower Gaussian used to generate the DOG filter and thus corresponds to the center of neighboring filters.}%
\lFig{dog}%
\end{figure}%
As suggested by wavelet theory, we set $\sigma(1)=.5$ for the mother wavelet that defines scale $1$, so that at scale  $s$ (and up to maximal scale $s_\mathtt{max}$), ${\sigma(s)}=2^{s-2}$. At scale  $s$, the activities of the ganglion cells are calculated over the down-sampled grid $\mathcal{D}_s =\{ (x,y)= (2^{s-1} .i,2^{s-1} .j)\}$ where $(i,j)$ are natural numbers. The multi-scale representation therefore constitutes a \textit{dictionary} of filters $\mathcal{D}=\bigcup_{1 \leq s \leq s_\mathtt{max}}\mathcal{D}_s$ placed on progressively more widely spaced grids, hence the name of a \textit{pyramid transform}, the lowest scale being the base. The total number of neurons is thus proportional to the number of pixels by a factor of  $\chi =1+(1/2)^2+\ldots+(1/2)^{(2.s_\mathtt{max})}=\sum_i (1/2)^{2.(i-1)}= \frac{1-(1/2)^{s_\mathtt{max}}}{1-(1/2)^2}$ that is approximately $\chi=4/3$. Instead of differentiating ON or OFF cells (so that the number of neurons is doubled), we will consider for simplicity and because it is exactly equivalent that each neuron $i$ is assigned a polarity $p_i$ which is either $+1$ or $-1$, so that the coefficients are rectified (\ie $|C_{i}|=p_i. C_{i}$). \\
These filters resemble the receptive fields that can be observed in the biological retina \cite{Enroth-Cugell66}. However, they may be difficult to implement since first they decay slowly, but also it is necessary when sampling filters on the rectangular grid to correct them in order to avoid artifacts and to assure that the sum of the filters' coefficients is zero. From the Calder\'{o}n formula, the wavelet transform may be inverted and in our case the image may thus be reconstructed by the coefficients' values simply by
\[
I_\mathrm{rec} = K. \sum\nolimits_{i\in \mathcal{D}} C_i. \phi_{i}
\]%
where $K$ is a constant dependent on the filters which for simplicity is set to 1. %
In this architecture, the filters form an approximate \textit{orthogonal wavelet transform} \cite{Mallat98} of the  image, \ie the responses of different fibers are uncorrelated (that is $ <\phi_{i},\phi_{j}>=0$ for $i \neq j$). Note that it can be proved that this relation is here only approximate and that the reconstructed image is blurred. This blur is characterized by a "point spread function"\footnote{Similarly as in optics, this is the response of the whole system (coding and decoding) to an impulse, here to  the image of a single pixel at luminance $1$. From the linearity of the transform, it proves the assertion.} of the response :
$$
I_\mathrm{rec} = I\ast \mathtt{PSF} \mbox{ with }
 \mathtt{PSF}=\sum_{1\leq s \leq s_\mathtt{max}} \frac{1}{\sigma_s^2} < \phi_{\sigma_s} , \phi_{\sigma_s}>
$$
This linear layer therefore exhibits two problems : first, the reconstruction is approximate and second, its implementation may be computationally slow because the size of the filters can become very large. A common solution is to use a Laplacian pyramid as defined by Burt \textit{et al.} \cite{Burt83laplacian}. This transform uses a bijective down- and up-sampling scheme to simulate the convolution by the larger filters by using recursive filtering and a sub-sampling algorithm. This transform is still linear and the image may similarly be reconstructed as a linear combination of the filters by using the coefficients of the pyramidal transform. Moreover, this transform is computationally more tractable and since it is orthogonal, the reconstruction of the image is perfect. Here, we will at first use and compare both methods.\\%
Finally, during this linear stage, a static image is transformed by the linear model into a multi-scale representation of contrasts (see Fig. \ref{fig:archi_c}). These analog values are theoretically sufficient to reconstruct the image and under the assumption of orthogonality, this representation which is often used in image processing algorithms, provides a compact code of the image, \ie one in which the number of significant coefficients is relatively small.%
\subsection{Analog to latency coding: ranking multiscale contrasts}
\begin{figure}[h]
\includegraphics[width=\columnwidth]{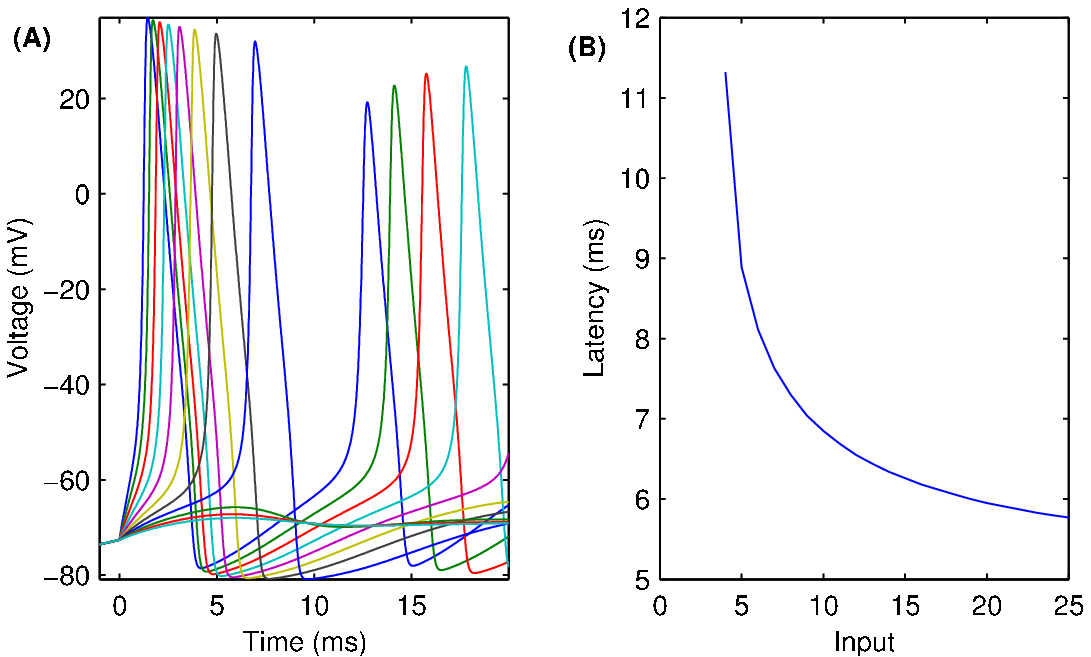}
\caption{Analog to latency coding. (A) We simulated a detailed Hodgkin-Huxley model neuron with steps of increasing activity. The neurons generated spikes (when reaching a threshold potential of $\sim$\unit[53]{ms}) with a frequency proportional to the activity which is revealed by the inverse latency between 2 spikes. (B) Similarly, there is a direct relation between the input (the rectified contrast in our model retina) and the latency. Above a threshold,  the spike's latency is increasingly shorter for progressively higher input activities.}
\lFig{HH}
\end{figure}
When presenting an image at an initial time, each neuron of the model integrates the analog contrast information at its soma until it eventually reaches a threshold: it then emits a spike. As in most models of neuronal integration, we will simply assume that the stronger the activation, the earlier the cell will reach the threshold. Such behavior happens in most biological neurons and can be be implemented by both detailed (see \seeFig{HH}) and more simple models such as the Integrate-and-Fire neuron \cite{Lapicque07}. Finally, the spike then propagates along the axon  and the neuron's activity is reset. Classically, this generates a pattern of spikes whose instantaneous frequency may constitute the image's code. But the code may also be equivalently carried by the exact spiking time (or {\em latency}) of the first spike. We may thus consider only the first spike, so that the code exactly consists of this latency for each of the different fibers $i$ and which is inversely proportional to the neuron's excitation current, \ie to the corrected activity. This algorithm defines a coding scheme that transforms an analog matrix pyramid into a spike 'wave front' that travels along the optic nerve.\\%
Using this framework, the coefficients are emitted and transmitted in order, starting with the highest rectified contrast.%
If we know exactly the corresponding contrast values when trying to decode the spike wave, we may reconstruct progressively the image by
\[
I_\mathrm{rec}(t) = \sum\nolimits_{r=1\ldots t} C_{o(r)}. \phi_{o(r)}
\]%
where $t$ is the corresponding discrete time corresponding to the count of fired spikes (\ie their rank) that we use for the reconstruction and $o(r)$ is the address of the neuron of rank $r$.  In fact, if we assume that the filters are orthonormal and from Pythagoras' theorem, since $o$ is a permutation of the addresses of neurons, the squared error $\mathrm{SE}(t)$ at time $t$ is simply:
\begin{eqnarray}
\| I_\mathrm{rec}(t) - I \|^2 &=& \| \sum\nolimits_{r=1\ldots t} C_{o(r)}. \phi_{o(r)} - \sum\nolimits_{i} C_{i}. \phi_{i} \|^2 \nonumber \\
 &=& \| \sum\nolimits_{r=1\ldots t} C_{o(r)}. \phi_{o(r)} \nonumber \\
&&- \sum\nolimits_{r=1\ldots r_{max}} C_{o(r)}. \phi_{o(r)}\|^2 \nonumber \\
 &=& \| \sum\nolimits_{r=t+1\ldots r_{max}} C_{o(r)}. \phi_{o(r)}\|^2 \nonumber \\
\mathrm{SE}(t) &=& \sum\nolimits_{r=t+1\ldots r_{max}} |C_{o(r)}|^2 \lEq{error}
\end{eqnarray}
where $r_{max}$ is the final time (and therefore corresponds to the total number of rectified coefficients). From \seeEq{error}, this strategy of coefficient propagation corresponds thus to a "greedy" minimization of the MSE at each step of the algorithm.  This also leads to the convergence of  $I_\mathrm{rec}(t)$ toward  $I_\mathtt{rec}$ (and therefore to $I$ for the Laplacian Pyramid), leading to a progressive compact coding of the image (see \seeFig{mse}).%
\begin{figure}
\includegraphics[width=\columnwidth]{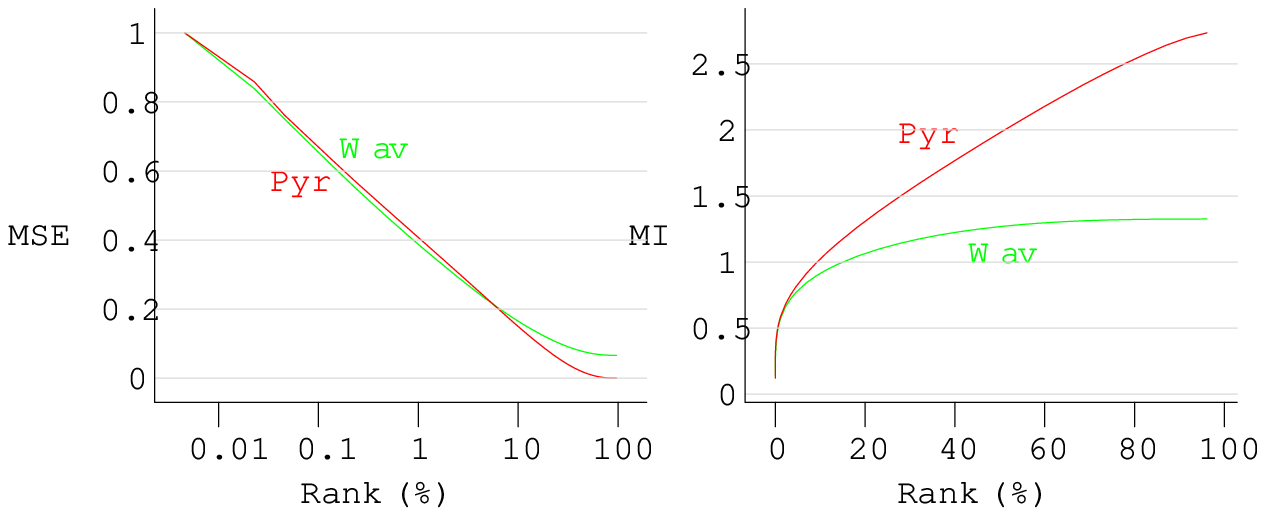}
\caption{Progressive reconstruction of the image from the ranked linear multi-scale representation. We plotted the Mean-Squared Error (MSE, logarithmic abscissa scale) and the Mutual Information (MI) of the progressive reconstruction using the \textit{exact} transforms' coefficients ranked from the highest energy to the lowest. We compared the method using a discrete wavelet transform (Wav) and a Laplacian Pyramid scheme (Pyr). It should be noted that this latter method provides at the end of the propagation an exact reconstruction of the image and the Mutual Information therefore converges to the mean entropy of the images in the database.}
\lFig{mse}
\end{figure}
\subsection{Interpreting ranks as activities : a cooperation in time} 
But, how can this information be encoded and decoded using only one spike per axon? Van Rullen and Thorpe \cite{Vanrullen01} have shown that these values observed regularities across natural images as they were ordered from the largest to the lowest. A solution is therefore to use the mean analog value to form a \emph{Look-Up Table} (LUT) to decode the analog values back from their rank. They thus defined
\begin{equation}
\mathtt{LUT}(r) = E[| C_{o(r)} |] 
\lEq{lut1}
\end{equation} 
where $E$ denotes the average over a set of randomly chosen images from the database\footnote{Further averaging or learning schemes used here 200 randomly chosen images.}. Then, we can reconstruct the image from the spike list using
\[
\tilde{I}_\mathrm{rec}(t) = \sum\nolimits_{r=1\ldots t} \mathtt{LUT}(r).p_{o(r)}. \phi_{o(r)}
\]%
where $\tilde{I}_\mathrm{rec}(t)$ is the image reconstructed using the spikes rank at step $t$ and $p_{o(r)}$ the polarity of neuron corresponding to the $r^{th}$ spike. Using the orthogonality of the filters, the error $\mathrm{SE}_\mathrm{Lut}(t)$ is therefore using a same method as above (see \seeEq{error})
\begin{eqnarray}
\| \tilde{I}_\mathrm{rec}(t) - I \|^2 &=& \| (\tilde{I}_\mathrm{rec}(t) - I_\mathrm{rec}(t)) + (I_\mathrm{rec}(t)-I) \|^2 \nonumber \\
 &=& \| \sum\nolimits_{r=1\ldots t} (\mathtt{LUT}(r).p_{o(r)}-C_{o(r)}). \phi_{o(r)} \nonumber \\
 && + \sum\nolimits_{r=t+1\ldots r_{max}} C_{o(r)}. \phi_{o(r)}\|^2 \nonumber \\
\mathrm{SE}_\mathrm{Lut}(t)  &=& \sum_{r=1\ldots t} (\mathtt{LUT}(r)-|C_{o(r)}|)^2 + \mathrm{SE}(t) \lEq{error_lut}
\end{eqnarray} 
The reconstruction error is therefore the sum the quantization error added to the energy that has not yet been transmitted.  \seeEq{error_lut} also justifies the choice of the LUT as the mean (see \seeEq{lut1}) since it is the optimal estimator for the rectified coefficient as a function of its rank in the MSE metric.\\
\begin{figure}[h]
\includegraphics[height=4.5cm,width=\columnwidth]{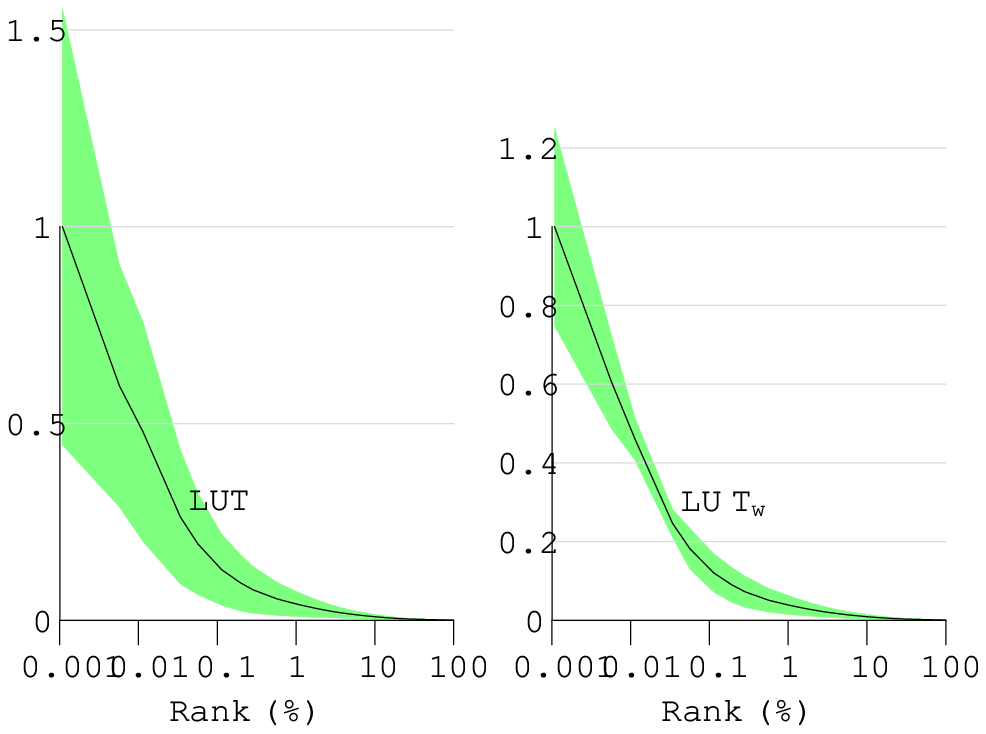}
\caption{\emph{Look-Up Tables} (LUTs) for decoding the analog value corresponding to a spike. These LUTs correspond respectively on the left to the average of the rectified contrast as a function of the rank and on the right to a mean of the same coefficients but weighted and normalized accordingly to provide a more robust regularity. The filled-in regions correspond to the variance of these measure and will be directly linked to the MSE of the reconstruction. The weighting process provides thus a better spike representation of the coefficients. }
\lFig{lut}
\end{figure}
Neurophysiological mechanisms for producing this decrease of the coefficients over time were discussed in \cite{Vanrullen01} and may involve a set of separate neurons ---interneurons--- using shunting inhibition \cite{Delorme03}. We propose here that since these "rank counter interneurons" could be tuned used an incremental adaptive rule with an on-line hebbian learning scheme. This rule takes the form of a stochastic algorithm so that after coding the $n^{th}$ image using $m^{(n)}$ as a modulation function,
$$
m^{(n+1)}(r) = (1-\mu^{(n)}) . m^{(n)}(r) + \mu^{(n)} . |C^{o(r)} |
\lEq{hebb_lut}
$$
where $t$ is as before the discrete time corresponding to the decomposition and $\mu^{(n)}$ (typically, $\mu^{(n)}=1/n$) the stochastic learning gain. In practice, this is exactly equivalent to the averaging algorithm (see \seeEq{lut1}) and thus leads to a similar reconstruction error.  A more realistic biological implementation would consider the limited receptive fields of these inter neurons, leading to a measure of the local rank, but although this would lead to interesting results, it is computationally very demanding. It should be noted that the rectified coefficient $|C^{o(r)}|$ has yet to be transmimtted and it can be computed over a longer time scale with for instance the mean frequency of firing. This mechanism provides thus an adaptive and complete mechanism for a temporal spike code using a temporal cooperation.\\%
This algorithm transmits rapidly and in a robust fashion the image by using the rank of the neurons, and it is in this task superior to other temporal spike coding strategies such as frequency coding \cite{Vanrullen01}. However, one may notice that the variance of the LUT is relatively high and especially for the first spikes, \ie for the most important spikes. Since we saw that the reconstruction error is proportional to this variance (\seeEq{error_lut}), a better strategy is to give more importance for the first ranks. In practice, we used a logarithmically decreasing "gain" function:
\[
m^{(n+1)}(r) = (1-\mu^{(n)}) . m^{(n)}(r) + \mu^{(n)} . |C^{o(r)}. (1-\log(\frac{r}{r_{max}}) |
\]
The resulting LUT yields less variance especially for the first spikes (see \seeFig{lut}) which leads to better convergence and  improved image transmission (see line 'LUT' in \seeFig{mse_lut_natu}).
\subsection{Distribution of the singularities in whitened natural images}
One may wonder why this regularity occurs in natural images. %
\begin{figure*}[!t]
\centering
\includegraphics[width=11cm, height=6cm]{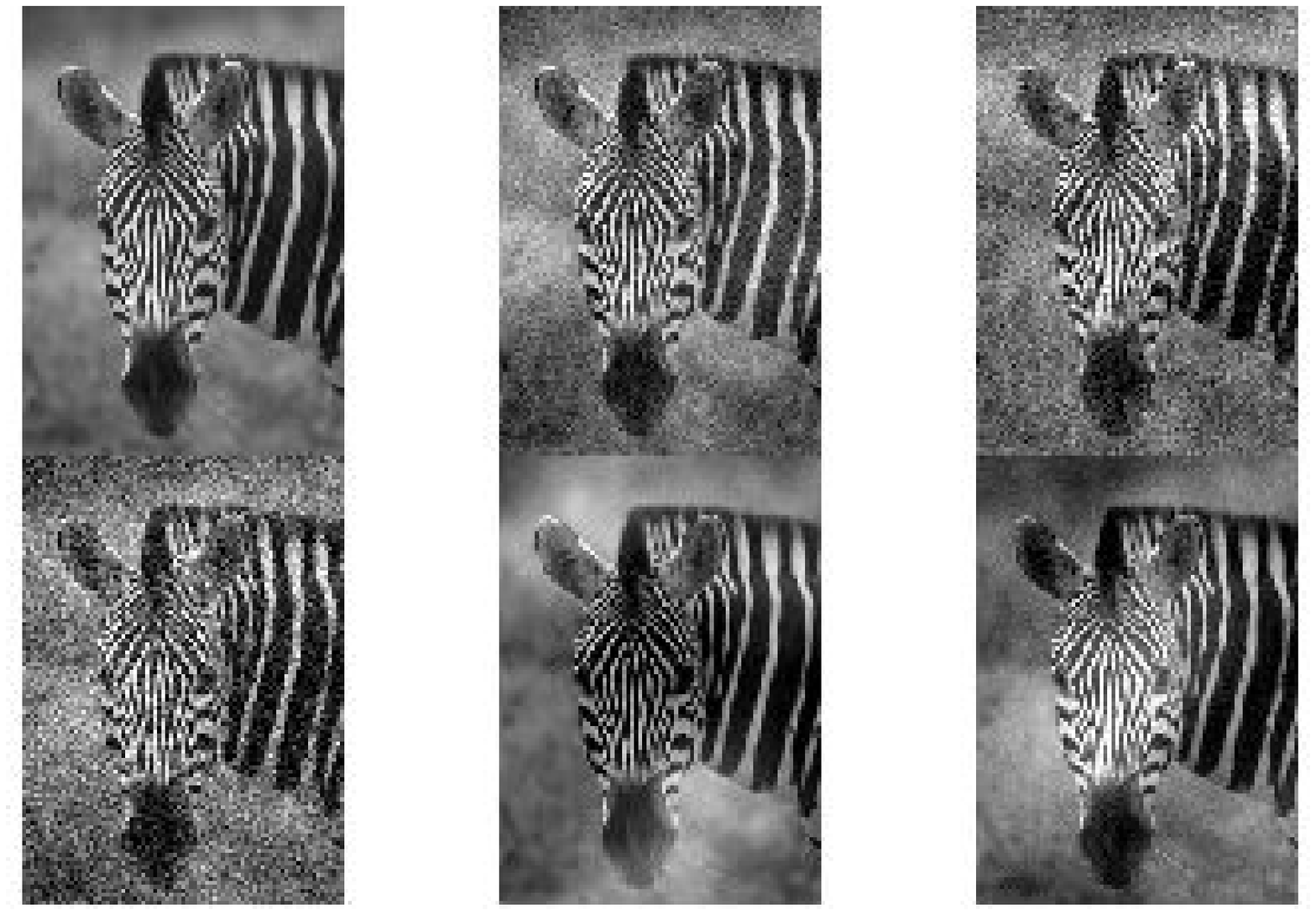}
\vskip -5cm
\includegraphics[width=11cm, height=6cm]{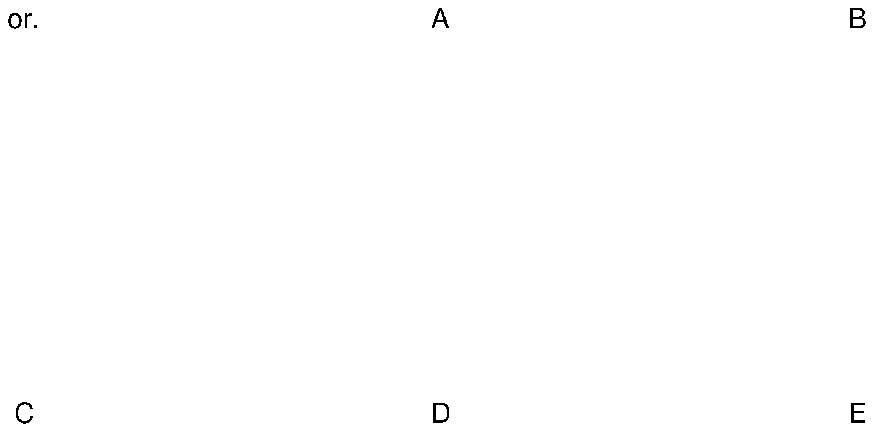}
\caption{Subjective distance between images. We added to an image {\bf (Or.)} different noises which are characterized by the envelope of their spectral energy to obtain 5 new noisy images {\bf (A)} to {\bf (E)} deviating from the original. Comparing the results for the Mean Squared Error and for the Weighted MSE, we may rank the quantitative distance of the noisy image compared to the original : for MSE closest to noisiest, we obtain A, B, E, C et D whereas WMSE provides D, E, A, B et C. This harmonized distance is more robust to changes at low frequencies and corresponds to a more subjective measure of the noise added to an image, \ie to a distance between images (see e.g. the zebra's ear).}
\lFig{dist_w}
\end{figure*}
In fact, when analyzing this regularity separately for the different scales, we proved that the coefficients of particular scales are not well tuned to the overall LUT (see \seeFig{why_natu}-Left) and that \textit{a priori} the coefficients corresponding to the lowest frequencies have a higher probability to be transmitted first whatever the image to be coded \cite{Perrinet02sparse}. From the "donut" shaped Fourier transform of the DOG filters, it is easy to see that there is a direct correspondence between the activities of the neurons at a given scale and the Fourier components of the image at a certain frequency. The mistuning of neurons at different scales thus corresponds in Fourier space to the shape of the mean power spectrum function. Over natural images, it is known to decrease in $\frac{1}{f^2}$ \cite{Field94}, a result from correlations between the luminances of neighboring pixels. We therefore applied a decorrelating kernel as defined and computed by Atick \cite{Atick92} to the input image. Note that this re-normalization according to the scale (or temporal frequency) leads to a different distribution of the Fourier components in the spatial frequency space: the image's power spectrum distribution is "spherized". At the same time, we can derive a new measure of the image reconstruction error based on this renormalization, that we denote as the \textit{Weighted Mean Squared Error} (or WMSE) and which leads to a new distance between images. The WMSE appears to be more correlated to a subjective measure of distances between images (see \seeFig{dist_w}), and since there is a non-uniform prior in the energy of coefficients as a function of temporal frequency, it corresponds in fact to the Mahalanobis distance \cite{Mahalanobis36} applied to our set of natural images. It removes some of the disadvantages of the MSE measurement, such as its dependence to a constant component and provides thus a new criteria for image reconstruction. \\
\begin{figure}[h]
\centering
\includegraphics[height=4cm,width=\columnwidth]{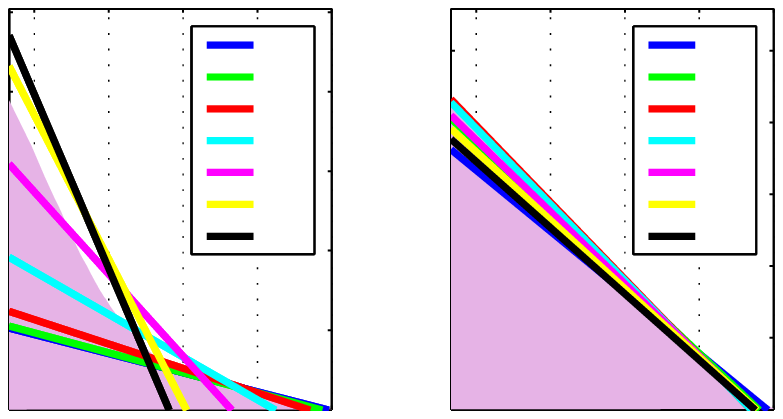}%
\vskip -4cm%
\includegraphics[height=4cm,width=\columnwidth]{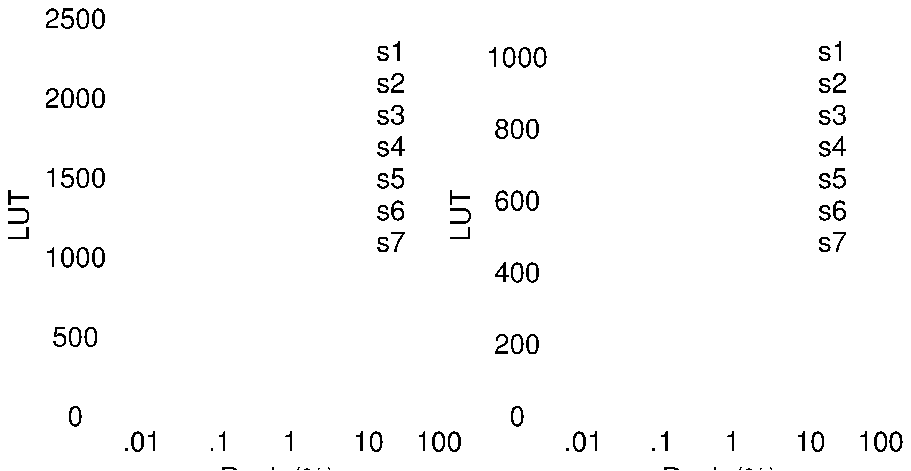}\\%
 \hspace*{.5cm}  Rank ($\%$) \hspace*{2.5cm} Rank ($\%$)
\caption{Optimization of the regularity of the wavelet coefficients with harmonized scales. The LUT is shown at the background in plain color as a function of the rank in $\%$ using a logarithmic scale on the abscissa. When separating the LUT for the different scales (from lowest to highest : $s1$ to $s7$ in the legend), one may observe that they correspond to similar regularities ---linked to the regular distribution of singularities at different scales--- but are mistuned (lower frequencies, as the $7^{th}$ scale 's7' are stronger and thus decrease more rapidly as a function of the overall rank). These regularities are therefore lost and mixed when ranking all scales together. By normalizing the different scales according to the statistics of natural images, the "vote" by the ranking process becomes "fair" and the LUTs for the different scales can be made to match. The resulting LUT for harmonized scales preserves the underlying regularity and information transmission is therefore more robust (see \seeEq{error_lut}): it represents  a more effective way to encode the analog value by the rank of a spike.}
\lFig{why_natu}
\end{figure}
Back to the linear model, this decorrelation process corresponds simply to a pre-process of filtering by the decorrelating kernel. In our model it would correspond to the introduction of a layer between the photo-receptors and the ganglion cells that mimics the behavior of the horizontal and bipolar cells in the retina and results in  modifications in the spatial frequency tuning of cells \cite{Enroth-Cugell66}. Since the scales are tuned, rectified coefficients follow a very regular linear decrease (see \seeFig{why_natu}) in the log-linear plot, starting at rank 1 to a value proportional to the mean energy in the image and ending at the final rank at zero. It suggests the existence of a relation of the  rectified contrast value as a function of the logarithm of the relative rank. We therefore used a similar averaging rule for the LUT (\seeEq{hebb_lut}) function which resulted in a more regular function. From \seeEq{error_lut}, and since this leads to less variance, we are thus assured that this regularity results in more effective information transmission.\\%
In fact, this regularity may be linked to the distribution of \textit{Lipschitz exponents} in natural images. They correspond to a measure of the order of the singularities which are present in the image, and that can be qualitatively ranked from the highest to the lowest Lipschitz exponents as : isolated dots, lines, edges, slopes, gradients until uniform surfaces. In our framework, since this multi-scale contrast representation gives a local measurement of the Lipschitz exponents in the image (see \cite{Mallat91}, \cite[p.513]{Mallat98}), we can qualitatively link the definition of its value $z$ with our model:
\begin{eqnarray}
\gamma(z) = - \frac{ d \log x(z) }{ d \log z } \nonumber
\end{eqnarray}
where here $z=\frac{r}{r_{max}}$ is the relative rank and  $x(z)=|C_{o(r)}|$ is the activity of neurons as a function of the rank\footnote{Since the energy of the image is equal to the sum of the squared coefficients, this constraint may be used to introduce a renormalization of $x$. In general, we verified at the start of each propagation that the total energy was equal to 1.}. When studying whitened natural images, we can observe that after normalization $x\sim-\log z$ and therefore  that the propagation scheme ranks singularities from their highest order to their lowest with a regular distribution.\\
The same was already observed for un-whitened ("raw") images after a certain rank (see \cite[p.513]{Mallat98}).  We may interpret the relative regularity of the distribution of Lipschitz exponents physically as (1) the whitening process removes the correlations between spatial frequencies due to size and depth of objects \cite{Alvarez99}, (2) then, the distribution of complexity of shapes and textures of objects in nature is regular. This last point is linked to the inherent properties of auto-similarity \cite{Turiel98,Turiel99} in images. In a generative model framework \cite{Olshausen98}, \ie in which we assume that all natural images may be generated by a probabilistic model, this result suggests that singularities are chosen with a characteristic probability: it is therefore an important feature of natural images corresponding to an important measure of the distribution of complexity in the image. It corresponds to a  high level parameter that can be used to generate the coefficients for the whole set of natural images whereas the ranked list $o$ of events' addresses would correspond to the realisation of this particular image\footnote{Inversely, a desired distribution of exponents can be generated with a particular modulation of coefficients as a function of the rank.}. This generative model approach justifies the use of the LUT in the algorithm since it corresponds to a physical interpretation of the visual input.\\%
\begin{figure}[h]
\includegraphics[width=.9\columnwidth]{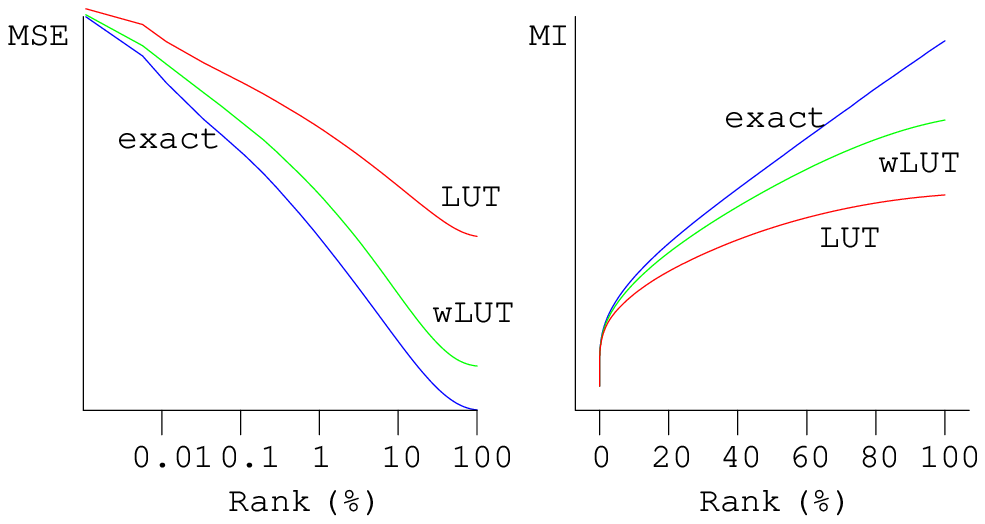}
\caption{Progressive reconstruction of the image from the spike list using rank-order coding. We plotted the Mean-Squared Error (MSE, logarithmic scale on the abscissa) and the Mutual Information (MI) using the different temporal spike codes described in the text. We compared the results of the propagation when knowing the coefficients (exact) with the method described in \seeEq{lut1} (LUT) which uses an optimized Look-Up-Table to "guess" the value of the coefficients from their rank. Finally we compared these strategies to the optimized method that uses the regularity found in natural images through the statistics of natural images (wLUT). The reconstruction from this latter method is close to the method with exact values and proves that the analog values may be transmitted using rank order coding. It therefore constitutes a compact spike code which provides a simple implementation of rank-order coding for static images. }
\lFig{mse_lut_natu}
\end{figure}
But as we now ranked the coefficients according to decreasing Lipschitz exponents, and since low frequencies seem to provide \textit{a priori} information that is physically closer and thus more useful for rapid categorization, we may still want to propagate the coefficients according to their energy, \ie propagate the lowest frequencies first. In fact, since the use of normalized filters already provides this feature, this was accomplished in the model of Van Rullen and Thorpe by the fusion of (1) the use of the regularity as a function of the rank of coefficients to code coefficients with analog values with spikes (2) the ranking of the information carried by neurons according to their importance for the progressive propagation of the information.  However, we have shown that this is incompatible with the regularity in natural images and we will overcome this problem by separating these two ranking processes. As is implemented in the retina by the differentiation between the Magno- and Parvo- cellular pathways, low and high spatial frequency bands show different mean latencies, the neurons from the Magno-cellular pathway being significantly faster. Similarly, we can still rank the weighted coefficients (in WMSE metric) to produce a highly regular LUT (as in \seeFig{why_natu}-Right), hence a better transmission of the coefficients but now rank the propagation of the coefficients according to their energy (in MSE metric) so as to choose the order in which the spikes are emitted (therefore using a similar algorithm as the first scheme). In a probabilistic inferential model, this will correspond to the inclusion of a gain for low frequencies when in the context of rapidly detecting an animal. This scheme results in improved transmission of the image with a result that is close to the reconstruction using the exact coefficients (see \seeFig{mse_lut_natu}, line wLUT).\\
As a conclusion for this model, we have provided a general scheme for temporal spike coding using the relative rank and using the statistics of natural images. Practically, the scheme uses two parallel sorting mechanisms, one based on the regularity of the distribution of Lipschitz exponents and the other based on the progressive transmission of the parts of the image starting with the most informative. Together, they provide an algorithm that can efficiently decode the analog values corresponding to each spike using only the rank order information. This proves that this strategy can build a complete and efficient code from the retina (analog to spike coding) which can be decoded (spike to analog coding) using solely a temporal cooperation  between the successive neurons that fire, \ie a \emph{rank-order coding} scheme \cite{Thorpe96}, which provides a compact \textit{temporal spike code} in the retina.
\section{Non-orthogonal representations, toward a sparse temporal spike code}
\subsection{Orthogonal vs. non-orthogonal representations}
The condition on the filters for a perfect reconstruction ---\ie the orthogonality of the dictionary used to represent the image--- is a strong constraint on the architecture and is achieved only approximately with the model presented in \cite{Vanrullen01}, resulting in a small information loss. Moreover, in the biological retina, the architecture is not dyadic and real neighboring neurons can often have correlated responses and the previous model would result in a redundant representation. This condition is therefore too restrictive in order to build a biologically reasonable model of the retina where the response of neurons depend upon the activity of neighboring cells \cite{Meister01}, that is where they may cooperate spatially. Such restrictions would be even more problematic if we wanted to apply the same spike coding algorithm to cortical models as the primary visual cortex where the interdependence is even stronger. \\%
In fact, in order to code the image with a linear generative model, we may want to use an \emph{over-complete representation} of the image, \ie one which the number of filters is far greater then in the previous model. Such representation result in a \emph{sparse code}, \ie one in which the absolute values of the underlying linear generative model decrease rapidly \cite{Olshausen98}. But mathematically optimizing the linear generative model  leads to a combinatorial explosion of the freedom of choice  of the filters and of their corresponding coefficient values (it is a \emph{NP-hard} problem \cite{Mallat98}).%
\subsection{Spike coding using a Matching Pursuit : adding spatial cooperation to rank-order coding neurons}
\begin{figure*}[ht]
\centerline{\includegraphics[width=12cm, height=6cm]{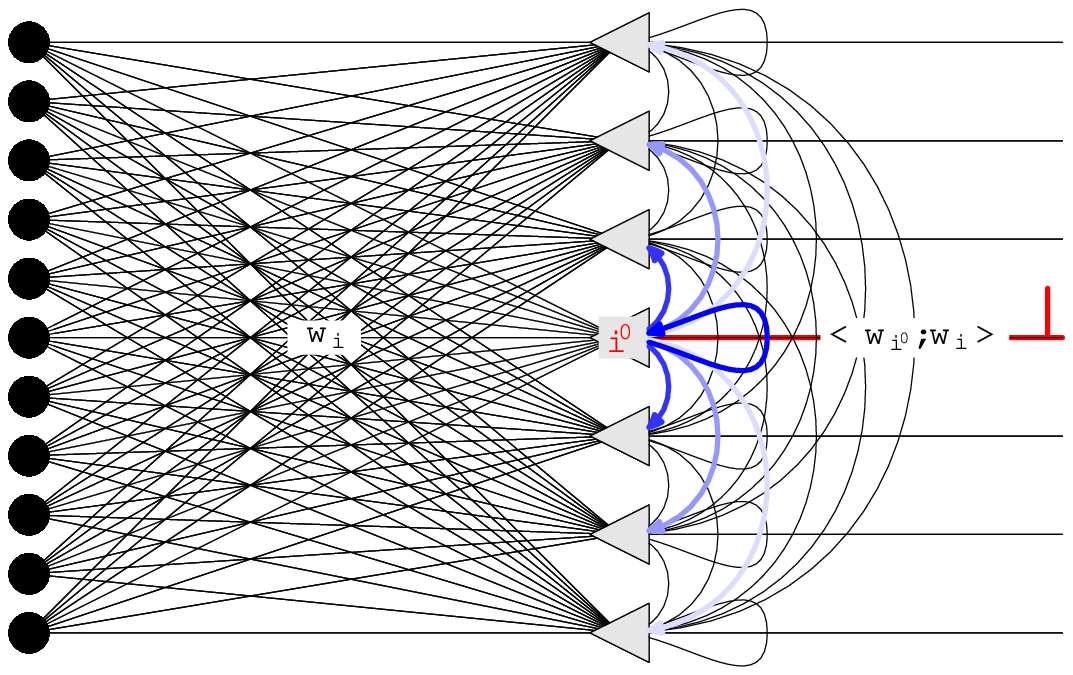}}
\caption{Principle of spike coding using a matching pursuit scheme. We represented a layer of neurons $i$ sharing some similar inputs on their synapses (black dots) according to given synaptic weights $\vec{w}_i$. The idea is to select at a time step $t$ the neuron $i^t$ corresponding to the maximal activity and to elicit a spike along its axon. Using a matching pursuit scheme, we may then directly account for the correlation between filters by subtracting from the other neurons an amount of activition proportional to the correlation $<\vec{w}_{i^t},\vec{w}_i>$ between their weight vector and the weight vector of the chosen neuron. The algorithm is then resumed at the next time step with the choice of a new winning neuron. This spatial cooperation ---the firing of one neuron is accounted in correlated neurons--- is then recursively repeated to the neuron corresponding to the maximal updated activity and until the maximal activity is less than a given threshold. The spatio-temporal spike pattern will therefore represent the input signal and may be reconstructed by a simple linear rule (see \seeEq{mp_rec}).}
\label{fig:mp}
\end{figure*}

Another strategy is to use a \emph{Matching Pursuit} (MP) \cite[pp.412--9]{Mallat98} algorithm, which is derived from a statistical estimation algorithm that has also been extended to wavelet theory \cite{Mallat93}. The idea is that we have to account for the correlations between filters and we therefore need to build up lateral interactions to cancel the correlation whenever a filter is selected. The MP algorithm decomposes the image over a large arbitrary dictionary $\mathcal{D}$ by iteratively choosing the best match and then removing the orthogonal projection of this match. \\
In this progressive scheme, let us first set the initial image $I^0=I$ and activities  $C^0_{i}=C_{i}$ at the initial time $t=0$. Then, we determine the first neuron in the layer to fire as the one with the highest activation (see \seeFig{HH}):
\[
{i^0}=\mbox{ArgMax}_{i} (| C^0_{i} |)
\]
For this index $i^0$ (the \textit{address} of the neuron), we define the corresponding extremal contrast value $C^{0}_{i^0}$. Since we have found the best match in the sense of the projection of the image on the dictionary, we can subtract the projection of this match $\phi_{i^0}$ (with norm $N_{i^0}$) to $I^0$ in order to define a first residual $I^1$ at time  $t=1$:
\[
I^1 = I^0 - \frac {<I^0  , \phi_{i^0}>}{\| \phi_{i^0} \|^2 } . \phi_{i^0} 
      = I^0 -  \frac {C^{0}_{i^0}}{{N_{i^0}}^2} . \phi_{i^0}
\]
The activity becomes at the same at time  $t=1$:
\[
C^1_{i}= <I^1 , \phi_{i}>  =
 C^0_{i} - \frac {C^{0}_{i^0}}{{N_{i^0}}^2} . <\phi_{i^0},\phi_{i}>
\]
This defines a spatial cooperation of the winning neuron to the correlated neighboring neurons. Note that in a neurophysiological model, we do not need to update the image's intensities (backward propagation) because we can directly modify the activity in the adjacent neurons using a lateral propagation. We therefore associate to each spike a lateral interaction $<\phi_{i^{0}} , \phi_{i}>$ which accounts for the selected spike. Note in particular that $C^{1}_{i^0}=0$, \ie the activity corresponding to the best match at time $0$ is totally cancelled at time $1$ (see \seeFig{mp}). Iterating these steps, we may repeat this algorithm to obtain successive residual activities at the discrete times $t$ defined by the exact spiking times.  The progressive reconstruction is then simply at time step $T$:
\[
I_\mathrm{rec}(T)=\sum\nolimits_{t=0,\ldots ,T}  \frac {C^{t}_{i^t}}{{N_{i^t}}^2} . \phi_{i^t}
\]%
This algorithm is exactly equivalent to MP for normalized filters ($N_i=1$) and presents the same computational complexity and properties \cite[pp.412--9]{Mallat98}. In particular, the  convergence of the reconstruction is guaranteed \cite[p.414]{Mallat98} under the condition that the dictionary is at least complete\footnote{In fact, an over-complete dictionary may be incomplete, \ie when the space generated by all linear combinations of the dictionary's vectors does not recover the input space. But, in our case these the chosen dictionaries at least include a complete basis.}. It is important to note that since we subtract the projection, the residual image is orthogonal to the winning filter, a property which produces a similar relation for the MSE as for \seeEq{error} although filters in the dictionary are here generally not orthogonal. The MP  scheme thus provides a similar representation as before but avoids redundancies between the events representing the information. With an over-complete dictionary, this coding strategy provides a sparse representation of the signal: the number of coefficients needed to describe the image is much lower than the dimension of the input space.\\
As with the wavelet transform, it may be similarly translated to a spike coding scheme by associating to each step the firing of a spike and by evaluating a LUT, so that the coding algorithm is simply for $t \ge 0$,
\[
\left\{
\begin{array}{l}
{i^{t}} = \mbox{ArgMax}_{i \in \mathcal{D}} (|C^{t}_{i}|)\\
C^{t+1}_{i} =  C^{t}_{i} - p^t . m^{t}. \frac {<\phi_{i^{t}} , \phi_{i}>}{{N_{i^t}}^2}
\end{array}
\right. 
\]
with $ m^{t} = E[ |C^{t}_{i^t} | ]$ and $p^t$ is the sign of $C^{t}_{i^t} $ (\ie its ON or OFF polarity).  The reconstruction is then simply
\begin{equation}
I_\mathrm{rec}(T)=\sum\nolimits_{t=0,\ldots ,T} p^t.m^{t}.\frac {\phi_{i^t}}{{N_{i^t}}^2}
\lEq{mp_rec}
\end{equation} 
 In comparison with a wavelet decomposition, since the choice of the $n^{th}$ filter depends on the spike list for the previous times, this transform is non-linear. In particular, it is not possible to directly use \seeEq{error_lut} since the residual is not necessarily orthogonal to the inhibition. Rather, the quantization error is added to the residual image and may therefore be coded in following spikes : the propagation is adaptive and the quantization error does not necessarily add up monotonously as in \seeEq{error_lut}.%
\subsection{Rank Order Coding with Matching Pursuit in the retina}
To compare this algorithm with the model of Van Rullen and Thorpe \cite{Vanrullen01}, we kept at first the same dyadic architecture and observed the behavior of the values for the absolute coefficients as a function of the rank of propagation for different natural images drawn from a database of indoor and outdoor scenes. As in the previous model, we observed regularities across natural images that were again sufficiently stable to allow the use of a Look-Up Table (LUT) in order to decode the analog value by its rank. In particular, we observed the same regularity of singularity distributions when whitening the image by appropriately tuning the norm of the filters as a function of their scale.\\
\begin{figure}[h]
\includegraphics[width=8cm, height=4cm]{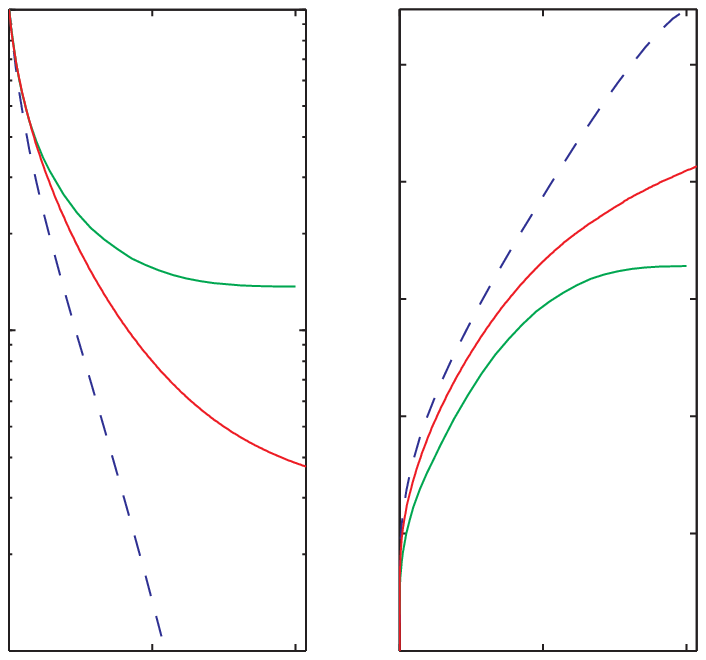}
\vskip -4cm
\includegraphics[width=8cm, height=4cm]{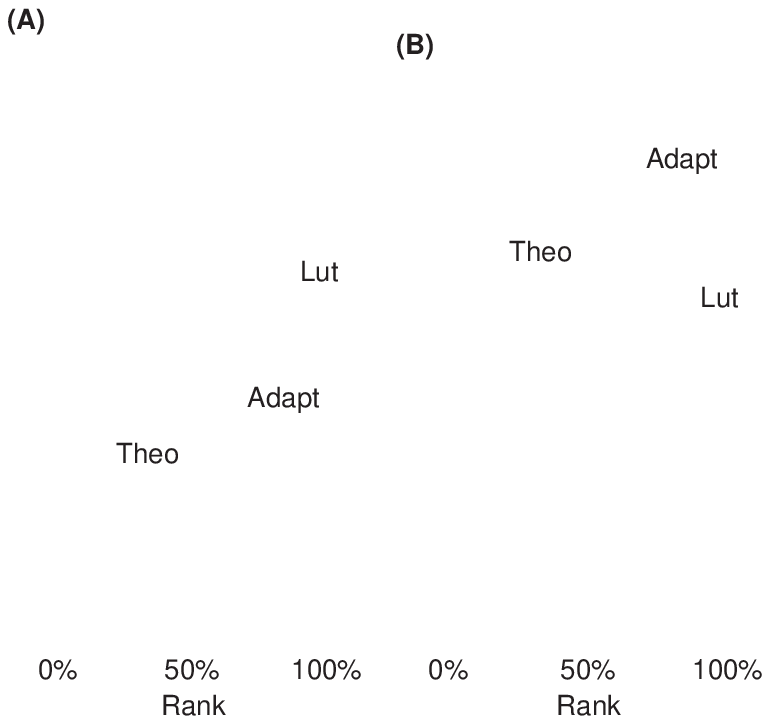}
\caption{Rank Order Coding with Matching Pursuit in the model retina. For the architecture defined in \cite{Vanrullen01} we calculated {\bf (A)} Mean Squared Error and {\bf (B)} Mutual Information of the reconstruction as a function of the relative rank (the percentage of the number of spikes fired to the total number of neurons) for the different coding strategies, comparing  (Theo) the theoretical reconstruction from the orthogonal wavelet coefficients, (Lut) the Orthogonal wavelet coding using a Look-Up Table as in \cite{Vanrullen01} and (Adapt) the Matching Pursuit with on-line learning (the image database consisting of 100 images to learn the modulation function and 100 images to measure the reconstruction  error). The adaptability of the MP algorithm enhances the transmission of the image and proves that the relative order of the action potentials could be used as a code in the optic nerve.}
\label{fig:comp_dya}
\end{figure}
Using the mean absolute coefficients as a LUT, we thus built a mechanism of reconstruction from the spike list, but as opposed to \cite{Vanrullen01}, this algorithm is adaptive and therefore  the error may be compensated dynamically, as opposed to \seeEq{error_lut}. Though filters are almost orthogonal (so that lateral interactions between filters ---\ie their correlation--- are relatively low) the MP algorithm introduces a gain in both the sparsity of the coefficients and in the reconstruction quality (\seeFig{comp_dya}).%
\subsection{Is the spike representation over-complete in the retina?}
\begin{figure}
\centering
\includegraphics[height=4cm,width=\columnwidth]{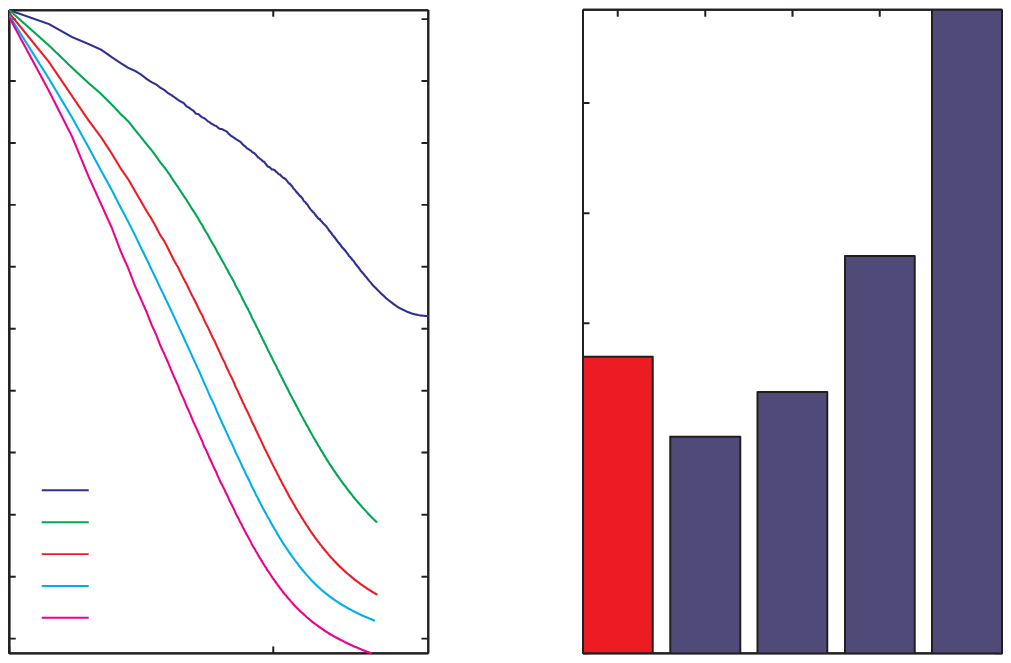}%
\vskip -4cm%
\includegraphics[height=4cm,width=\columnwidth]{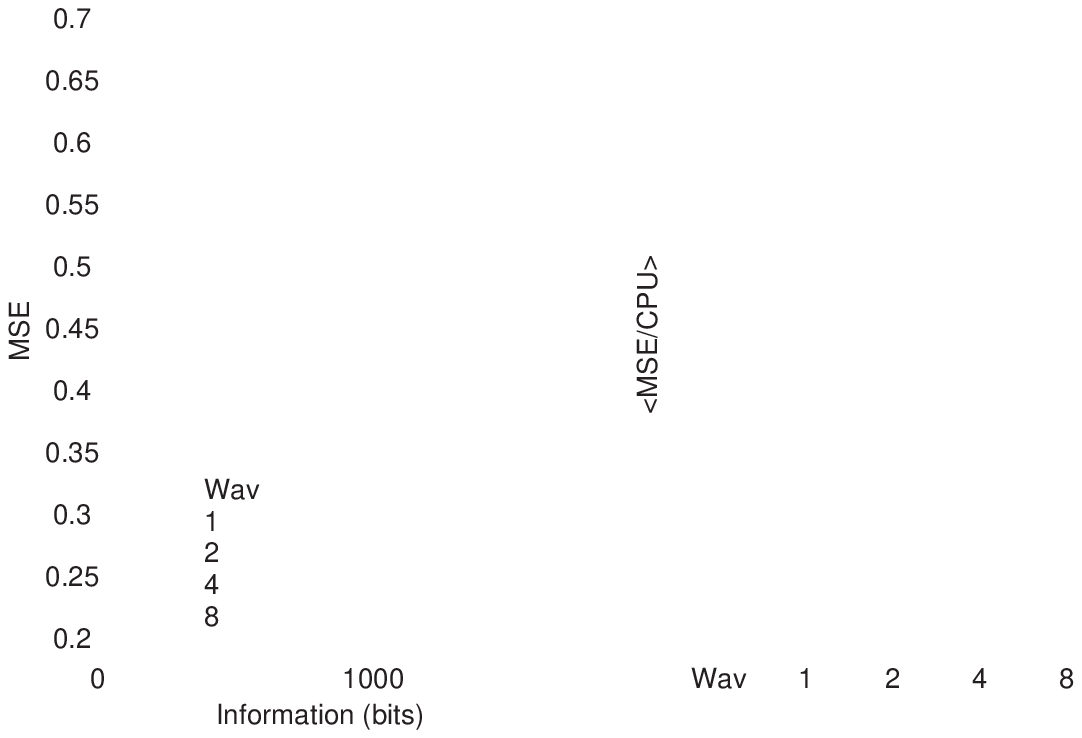}\\%
\caption{Is the spike representation over-complete in the retina? (Left) We compared the progressive transmission of information for different degrees of over-completeness in the retina by plotting the average MSE of the residual as a function of the information to code the spike list (in logarithmic scale, propagation up to $12.5\%$ of the relative rank for clarity). The set of neurons used rotation symmetric Mexican hat filters, with scales from layer to layer growing as $\rho=\{ 2,\sqrt{2},\sqrt[4]{2},\sqrt[8]{2} \}$ (and denoted on the legend respectively as 1, 2, 4 and 8). As a comparison we plotted the method used in the first part of the text (line 'Wav'). As a function of rank, the MSE decreases more rapidly for increasing degrees of over-completeness. (Right) But if we plot the trade-off of MSE with CPU usage as a function of the over-completeness, we find that for the same amount of information the adaptive dyadic strategy is  optimal.}
\label{fig:oc}
\end{figure}

 But now, considering the same spike coding scheme, we may ask whether an increase in the number of filters used to describe the image can enhance the representation, \ie if there would be an advantage to using an over-complete spike representation in the retina. We thus compared the sparse spike code for different degrees of over-completeness by choosing alternative progressions to the standard dyadic scale. The filters are thus defined as above, but the image pyramid now includes respectively $\{1,2,4,8\}$ scales \textit{per octave}, \ie the scale level characteristic variances now grow as $\sigma(s)=\sigma(1).\rho^s$ where $s$ is the scale index and $\rho=\{2,\sqrt{2},\sqrt[4]{2},\sqrt[8]{2}\}$.\\
These experiments proved that as the number of neurons increased, the coefficients decreased more rapidly as a function of the relative rank and also the MSE. This behavior is understandable, because choosing a higher number of filters allows the construction of a more fine grained multi-scale representation of the image. In fact, the number of neurons is multiplied by a factor $\chi =1+(1/\rho)^2+\ldots+(1/\rho)^{(2*s_\mathtt{max})}=\sum_i (1/\rho)^{2*(i-1)}= \frac{1-(1/\rho)^{s_\mathtt{max}}}{1-(1/\rho)^2}$ that is approximately $\chi =(1-\rho^{-2})^{-1}$.  This results in our different cases to an over-completeness of respectively $\{4/3, 2, 2 + \sqrt{2} \sim 3.41, 1/(1-1/\sqrt[4]{2})\sim 6.28 \} $. The information  (in bits) needed to code the address of each spike (position and scale) is thus $\log_2(n_{pixel})+\log_2(1-(1/\rho)^2)+1$  ($n_{pixel}$ being the number of pixels and one bit being allocated for the polarity). We may therefore compute the performance of the coding scheme in terms of the mean decrease in MSE as a function to the number of bits necessary to code the spike list  (see Fig.~\ref{fig:oc}-Left). However, the situation is different if we compare the trade-off between efficiency (MSE decrease) and the computational complexity (we assumed here that the CPU usage is proportional to the number of neurons). We obtained different results as a function of the degree of over-completeness  (see Fig.~\ref{fig:oc}-Right) and so conclude that under this constraint, the adaptive dyadic architecture would be optimal in the retina. \\
This appears to be mainly due to the nature of DOG filters (and to circularly symmetric wavelet filters in general) which to a certain extent overlap too much and does not capture any new information. In fact, the evolution of the retina is certainly constrained by its function, so that the argument may be reversed. First, the retina plays a key role in the visual pathways since it is the first processing layer : it is therefore very demanding in terms of robustness and the neurons are highly active. Moreover, the eyes are in wide range of living species are mobile elements which permit the active exploration of the visual environment. Thus, the number of neurons in the retina is presumably limited not only by the total energy it can devote but also by physical restrictions such as the size of the optic nerve. Since this number is limited (its over-completeness is limited), the representation may only use more general filters. Simulations of filter emergence in this framework (described in \cite{Perrinet03nc}) show that for a small number of filters, the optimal filters converge to contrast selective detectors (unpublished data). It is therefore interesting to study the case in the primary visual cortex where the situation is different: the information is there multiplexed and filters may be selective to different orientations.
\subsection{Over-complete representation in the primary visual cortex: Sparse spike coding}
Simple cells in V1 are known to exhibit a preference for oriented filters and we will here briefly present a model of over-complete representation using a dictionary of Gabor filters to compare the time course of temporal spike coding with coding in the retina. In comparison with the retina, the over-completeness in V1 is far greater (in humans the number of ganglion cells is of the order of one million whereas for V1 this number reaches at least $300$ million). In order to model the simple cells of V1, we used the spike coding scheme (as described in \cite{Perrinet02sparse}) with  a set of weight vectors $\psi_{j}$  defined as dilated, translated and sampled {\em Gabor} filters (see \cite[pp. 160]{Mallat98}). The scale grows geometrically with a factor $\rho = \sqrt[5]{2}$ (\ie 5 layers per octave) over $8$ octaves and the preferred orientation is circularly $0, \pi/4, \pi/2$ and $ 3\pi/4$.\\
\begin{figure}
\centering
\includegraphics[height=4cm,width=\columnwidth]{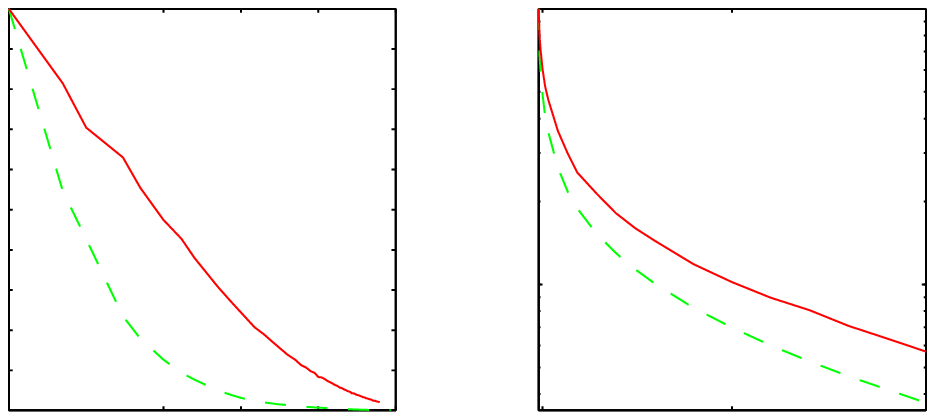}%
\vskip -4cm%
\includegraphics[height=4cm,width=\columnwidth]{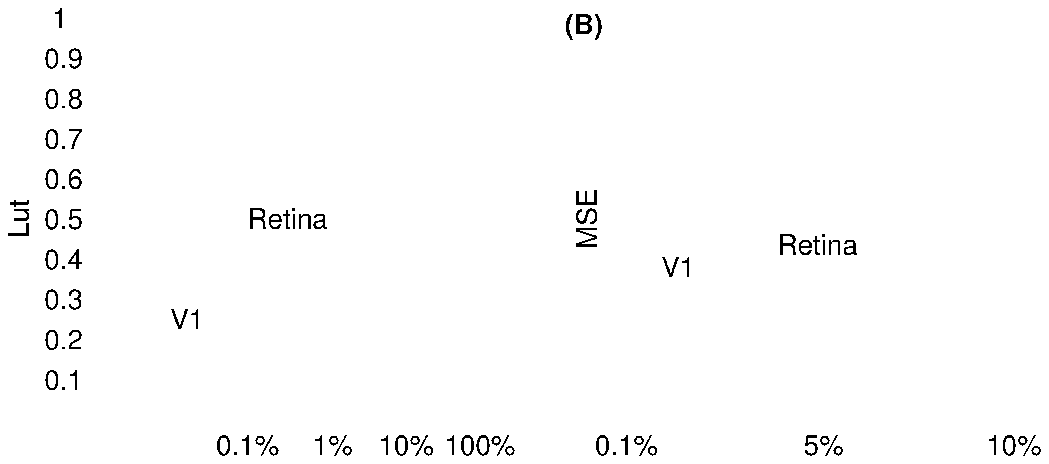}\\%
\vskip -.4cm%
\hskip 1cm {\tiny Rank} \hskip 3cm {\tiny Rank} 
\caption{Spike coding in the Retina and in V1. {\bf (A)} We computed recursively the LUT for the model adaptive dyadic retina and for the model of V1. In comparison with the retina, the coefficients decrease very rapidly for the model V1. {\bf (B)} MSE for the corresponding progressive image reconstruction (using logarithmic axis) defined by using this spike code. This proves that we defined an efficient visual code in V1 using an over-complete set of  Gabor filters and which leads to a model of a sparse spike code.}
\lFig{mp_lut}
\end{figure}
As described in \cite{Perrinet02sparse},  the LUT were generated in the same manner (see \seeFig{mp_lut}) and provided an efficient representation of static images. Moreover, the location of spike firings corresponded with the location of edges at different scales and order of singularity in the static images providing a dynamic 2 1/2 sketch of multi-scale contours. The resulting distribution of the coefficients is more kurtotic than in the retina (\ie it decreases more rapidly toward zero). Since the number of filters is higher,  the information rate ---\ie the information needed to code the address of one spike--- is now in this layer $\sim 16.1\unitfrac{bit}{spike}$. However, convergence is quicker so that this code may be compared with JPEG at high compression gains as shown in \cite{Perrinet02sparse}. This therefore defines a \textit{sparse spike coding} scheme in V1 (\seeFig{mp_lut}).\\
This brief description shows that the use of more complex filters may yield more efficient representations. Moreover, we proved that these filters could also be learnt using a simple hebbian learning scheme \cite{Perrinet03nc} leading to an adaptive scheme that can code natural images optimally. However, the optimal set of filters for V1 is still unknown, nor do we know the optimal degree of over-completeness for the dictionary (how many scales per octave? how many different orientations?). This open question needs first to solve the actual \textit{function} of V1 constrained by its structure. In fact, V1 ---as the most of the cortex--- is organized in 6-layered structure of elementary cortical columns which could provide a hint to the particular mechanisms underlying cortical processing. In particular \cite{Rao99predictive} suggested that these highly inter-dependant columns may implement a basic mechanism of inference, the whole system predicting future outcomes on the basis of the current input, the internal state and the expected gain predicting future states.
\section*{Conclusion : toward dynamic sparse spike coding}
We presented and analyzed here strategies of temporal spike coding that emphasize  the transient response of the neurons and showed how an event-based temporal code could be implemented using a rank-order scheme by the use of both temporal and spatial cooperation. In particular, we mathematically analyzed the model presented by Van Rullen and Thorpe \cite{Vanrullen01} which is based on an orthonormal wavelet representation and proposed strategies to improve the performance of the temporal cooperation used to code the information as a rapid spike wave. Moreover, by taking into account the statistics of natural images we have shown how regularities in the distribution in the order of the singularities in whitened natural images can be directly used to improve this spike-based code by providing two separate ranking strategies: one to precisely decode spikes as a function of their rank and a second that propagates the most useful information in the most efficient way.\\
We further extended this model to a model of sparse spike coding using arbitrary  representations by implementing lateral interactions which favor cooperations between neurons. We showed how this code is superior in the retina thanks to its adaptability, but also compared architectures with increasingly more over-complete representations. While a more precise sampling of the wavelet space provided more accurate representations, the adaptive dyadic transform appears to provide a near optimal compromise between efficiency and cost of computation. However, the use of more complex filters in higher cortical areas suggests that an over-complete dictionary can provide a computational gain. In particular, simulations with Gabor filters provided a temporal \textit{sparse spike coding} representation of the image which can be used to model V1. However, it seems still unclear if the chosen architecture for V1 is optimal in terms of the compromise between rapidity, precision and cost of computation.\\
This scheme provides a simple algorithm for image processing which proves to be very effective and that can be used in parallel algorithms such as {\sc Spikenet} \cite{Delorme99spikenet}. It shows specifically that in the model V1, the use of lateral interaction to reduce redundancies could provide a speed-up of the processing compared to an orthonormal feed-forward scheme. A particularly interesting extension of this scheme, would be first to introduce mechanisms described by Bullier \cite{Bullier01}: the spiking  information from one layer or one sub-layer can modify the sensitivity of neurons in another layer or sub-layer to account for the information already propagated. For instance, the rapid activity of neurons in the Magnocellular pathway could cooperate with neurons in the Parvocellular pathway by providing a coarse information.\\
At last, it would be interesting to extend this spatio-temporal cooperation to a spike code in the time domain. The matching pursuit scheme has already been used to build a video compression codec \cite{Neff97} and should be particularly efficient for processing video streams. These advances would thus introduce a precise paradigm of event-based computing mimicking the efficiency of temporal mechanisms in biological neurons.
\subsection*{Online simulations - reproducible research}
All scripts describing the models presented in the paper and reproducing the figures are available at :\\%
\url{ http://laurent.perrinet.free.fr/code/}

\end{document}